\documentclass[twocolumn,preprintnumbers,amsmath,amssymb,superscriptaddress,noshowpacs,showkeys]{revtex4-1}

\usepackage{graphicx}
\usepackage{dcolumn}
\usepackage{bm}
\usepackage[T2A]{fontenc}
\usepackage[utf8]{inputenc}
\usepackage[english]{babel}
\usepackage{multirow}
\usepackage{booktabs}
\usepackage{textcomp}
\usepackage{xcolor}
\usepackage{gensymb}

\usepackage{natbib}
\usepackage{float}
\usepackage{caption}
\usepackage{subcaption}
\usepackage[normalem]{ulem}

\usepackage{hyperref}
\hypersetup{
	pdftitle={ON THE ANGULAR ANISOTROPY OF THE DISTRIBUTION FUNCTION
    OF \\ RADIATING PARTICLES IN RELATIVISTIC JETS},
	pdfauthor={T.I.Khalilov, V.S.Beskin, V.I.Pariev}
}

\allowdisplaybreaks

\begin{document}

\begin{flushleft}
%УДК~524.354.4
\end{flushleft}

\title{ON THE ANGULAR ANISOTROPY OF THE DISTRIBUTION FUNCTION \\
OF RADIATING PARTICLES IN RELATIVISTIC JETS}

\author{\firstname{T.I.}~\surname{Khalilov}}
%\email{kalasxel@gmail.com}
\affiliation{Moscow Institute of Physics and Technology (National Research University), Institutskii per. 9, 141701 \\
Dolgoprudny, Moscow region, Russia}
\affiliation{P N Lebedev Physical Institute, Russian Academy of Sciences, Leninskii prosp. 53, 119991, Moscow, Russia}

\author{\firstname{V.S.}~\surname{Beskin}}
\email{beskin@lpi.ru}
\affiliation{P N Lebedev Physical Institute, Russian Academy of Sciences, Leninskii prosp. 53, 119991, Moscow, Russia}
\affiliation{Moscow Institute of Physics and Technology (National Research University), Institutskii per. 9, 141701 \\
Dolgoprudny, Moscow region, Russia}

\author{\firstname{V.I.}~\surname{Pariev}}
\affiliation{P N Lebedev Physical Institute, Russian Academy of Sciences, Leninskii prosp. 53, 119991, Moscow, Russia}

\date{\today}

\begin{abstract}
The observed power-law spectra of relativistic jets from active galactic nuclei clearly indicate a synchrotron mechanism of radiation by particles that similarly possess a power-law energy spectrum. However, the issue of their angular anisotropy has not been given sufficient attention until recently, although the example of the solar wind (where a strongly magnetized wind is realized in a similar way) shows the importance of taking this circumstance into account. In this paper, we study the evolution of an initially isotropic power-law spectrum of radiating particles as they propagate along expanding relativistic jets. It is shown that for relativistic flows in which the electric field plays a crucial role, the preservation of the first adiabatic invariant does not lead to a decrease in the pitch angles of radiating particles as they enter the region of weak magnetic fields. This is due to the drift nature of the particle motion.
\end{abstract}


\maketitle

\section*{Introduction}
\noindent

Relativistic jets emanating from active galactic nuclei are visible manifestations of their activity at an early stage of evolution~\cite{rev1, rev2}. 
At the same time, their hydrodynamic velocities correspond to Lorentz factor values $\Gamma$  of the order of 10--20, and the outflowing plasma retains these velocities at large distances from the core before it significantly slows down upon interacting with the intergalactic medium~\cite{Istomin, LPB}. There is no doubt that the observed radio emission from jets is related to synchrotron radiation from relativistic electrons. This is indicated by both the power-law spectrum of the observed radiation and its characteristic cutoff at low frequencies, explained by synchrotron self-absorption~\cite{LPG, Sokol}.

Unfortunately, the nature of the formation of the power-law spectrum of radiating particles is currently unclear~\cite{q0, q1, q2}. Therefore, in most studies, a power-law spectrum is directly postulated~\cite{pow1, pow2}. In this case, an isotropic distribution over pitch angles is assumed in a similar way. Plasma turbulence is usually referred to here, which will primarily lead to angular isotropization of the distribution function~\cite{turb, Sobacchi}. 

However, this approximation is valid only for high-density plasma in a weak magnetic field, that is, when the Alfven velocity is less than the speed of light and hence the velocity of particles. 
In this case, the effect of the pitch angle diffusion of the electrons interacting with the turbulent plasma will prevail over the effective acceleration, which leads to an isotropic distribution
of the pitch angles for the accelerated electrons~\cite{Lloyd}. For radiating particles in relativistic jets, the collisionless plasma approximation is valid in a strong regular magnetic field, for which a different behavior of the system may be true when a strongly anisotropic pitch angle distribution occurs. 

Indeed, according to numerical calculations (see, for example,~\cite{McTchB, Kosh}), 
turbulence in the central regions of relativistic jets does not play a decisive role in the dynamics of the outflowing plasma. In addition, as the example of the solar wind shows (which similarly represents a diverging strongly magnetized, however, non-relativistic flow with an certainly high level of turbulence), the electron temperature of longitudinal and transverse motion can differ several times~\cite{solwind0, solwind1, solwind2}. 
In other words, with a constantly operating mechanism leading to angular anisotropy, the resulting anisotropy of the distribution function can be significant even if the level of turbulence is high enough. Therefore, the question of angular anisotropy has been raised more than once when discussing the radio emission of relativistic jets. This anisotropy is supported by a number of observational facts~\cite{Clausen}.

Leaving the discussion of the effectiveness of the isotropization process in real astrophysical jets to the last part of this paper, we limit ourselves to analyzing the anisotropization mechanism itself; a separate paper will be devoted to determining the equilibrium angular distribution function. In other words, here we consider the model problem of the evolution of the angular and energy distribution of radiating particles as they freely propagate over long distances from the central engine in a regular magnetic field. Assuming, as is generally accepted, that the energy spectrum of the jet is power-law and the pitch angles are isotropic at the base of the jet, we show that for relativistic flows where the electric field is crucial, preserving the first adiabatic invariant does not lead to a decrease in the pitch angles of radiating particles as they enter the region of weak magnetic fields. This is due to the drift nature of the particle motion. 
At the same time, despite the fact that the index of the power-law spectrum of the radiating particles does not change, the distribution function itself becomes essentially two-dimensional, i.e., it depends not only on the energy of the particles, but also on their angular distribution.

The paper is organized as follows. In the first part, the formulation of the problem is discussed in detail, within which we study the evolution of the distribution function of radiating particles $f(\gamma, \mu, r)$ 
as a function of their energies (Lorentz factors $\gamma$) and pitch angles $\chi$ ($\mu=\cos\chi$). The second part is devoted to an exactly solvable problem (a monopole magnetic field in the absence of rotation), which demonstrates the adequacy of our approach to determining the distribution function $f(\gamma, \mu, r)$. 
Further, in the third part, a detailed study is carried out of the evolution of the distribution function of radiating particles as they propagate along a relativistic jet. This study takes into account both the rotation of the jet, which leads to the generation of an electric field, and the transition from a strongly magnetized to a weakly magnetized flow. Finally, in the last part, the main results obtained in this paper are summarized.

\section*{The problem}
\noindent

First, let us formulate a research question in which we aim to investigate the evolution of the distribution function of radiating particles. It is important to note that we only consider those reference frames in which the electric field is zero, as this is the basis of the theory of synchrotron radiation~\cite{pow1}. In this case, the particle distribution function has axial symmetry, i.e. it depends only on their energy and pitch angle $\chi$. 
Using now the generally accepted definition of the distribution function of radiating particles, denoted as $F({\bf p}, {\bf r})$ with normalization
$\int F({\bf p},{\bf r}) \, {\rm d}^{3}p = n_{\gamma}({\bf r})$, we arrive at a result for the ultrarelativistic particles under consideration ($p = m_{\rm e}c \gamma$)
\begin{equation}
{\rm d}n_{\gamma} = 2\pi \,  m_{\rm e}^{3} \, c^{3}\, \gamma^2 F(\gamma, \mu, {\bf r}) \, {\rm d}\gamma \,{\rm d}\mu,
\label{dn2}
\end{equation}
where $n_{\gamma}({\bf r})$ is the number density of particles, such that their initial energy exceeded the threshold of $m_{\rm e}\, c^{2}\gamma_{\rm min}$, and ${\rm d}n_{\gamma}$, respectively, is the number density of radiating particles, the Lorentz factors (energies) of which are enclosed in the range
$\left(\gamma, \gamma + {\rm d}\gamma\right)$, and the cosines of the pitch angle $\mu = \cos\chi$ are in the range $\left(\mu, \mu + {\rm d}\mu\right)$. 

It is important to note that here and below we assume averaging over the rotation phase $\varphi$, on which the distribution function is expected to depend in the presence of an electric field. However, due to the large-scale nature of electromagnetic fields, changes in energies and pitch angles within the phase can be neglected, since the typical size of the change in angle $\varphi$ is the Larmor radius, while the electric field under consideration (see below) varies on a scale of the light cylinder $R_{\rm L}$.

Now introducing for convenience the distribution function
$f(\gamma, \mu, {\bf r}) = 2 \pi \,  m_{\rm e}^{3} \, c^{3} \, \gamma^2 F(\gamma, \mu, {\bf r})$,
we obtain \mbox{${\rm d}n_{\gamma} =  f(\gamma, \mu, {\bf r}) \, {\rm d}\gamma \,{\rm d}\mu$} 
with normalization
\begin{equation}
\int  f(\gamma, \mu, {\bf r}) \, {\rm d}\gamma \, {\rm d}\mu = n_{\gamma}({\bf r}).
\label{norm} 
\end{equation}
Let us recall immediately that the relationship between number densities in moving and stationary (laboratory) reference frames in the general case is not given by relation
\begin{equation}
n_{\gamma}^{\rm lab} = \Gamma n_{\gamma}^{\rm com},
\label{nlab} 
\end{equation}
where $\Gamma$ represents the Lorentz factor associated with relative motion. However, it is fulfilled for an isotropic distribution of particles, which is assumed below for the comoving reference frame at the initial point of motion.

Moreover, we assume that the number density of radiating particles $n_{\gamma}^{\rm lab}({\bf r})$ 
is significantly less than the numerical density of the main (hydrodynamic) component $n_{\rm e}^{{\rm lab}}$; 
according to~\cite{Lenalambda, Lera}, this assumption is indeed valid 
($n_{\gamma}/n_{\rm e} \sim 10^{-3}$--$10^{-2}$). In this context, it is crucial to distinguish between the values of number density in the comoving and laboratory reference frames. In order to estimate the number density of the hydrodynamic component in the laboratory reference frame, we used a convenient method of normalization $n_{\rm e}^{\rm lab} = \lambda n_{\rm GJ}$, where
\begin{equation}
n_{\rm GJ} = \frac{\Omega B_{\rm p}}{2 \pi c e}
\label{GJ}
\end{equation}
is the so-called Goldreich-Julian number density. Here $B_{\rm p}$ represents a poloidal magnetic field, and $\Omega$ denotes the angular velocity at the jet axis.

Let us immediately discuss the numerical values of the parameters entered above. The estimated value of the characteristic angular velocity
\begin{equation}
\Omega \sim 10^{-6} \, {\rm s}^{-1} 
\label{Omega}
\end{equation}
is derived from a number of considerations. Thus, it follows from the analysis of the expression for the total energy release of the jet~\cite{ufn},
\begin{equation}
W_{\rm tot} \sim B_{\rm L}^2 R_{\rm L}^2 c,
\label{W_tot}
\end{equation}
where $R_{\rm L} = c/\Omega$ is the radius of the light cylinder, and $B_{\rm L} = B(R_{\rm L})$, which for most of the observed
jets is $W_{\rm tot} \sim 10^{44}$--$10^{46}$ erg s$^{-1}$. 
Numerical calculations~\cite{Zaman} indicate that the ratio of the light cylinder radius $R_{\rm L}$ to the gravitational radius $r_{\rm g} = 2GM / c^2$ of a black hole should be greater than approximately $10$. The same estimates were obtained from the analysis of the observed transition of the jet width from a parabolic to a conical shape~\cite{LenaOmega}. Finally, direct measurements of the rotation velocity of the jet in M87 give
\mbox{$\Omega = (0.9$--$1.3) \times 10^{-6}$ s$^{-1}$~\cite{Mertens}.}

Next, an estimate of the poloidal magnetic field at the black hole horizon 
$B_{\rm p}(r_{\rm g}) \sim (10^{3}$--$10^{4}$) G (the so-called Eddington magnetic
field, which follows from the equality of the magnetic field energy density and 
the plasma energy density for Eddington accretion rate~\cite{BZ, TM}) 
gives for the value of the poloidal magnetic field on the light cylinder 
$B_{\rm L} \sim (\Omega r_{\rm g}/c)^{2}B_{\rm p}(r_{\rm g})$
\begin{equation}
B_{\rm L} \sim (1-10^{2}) \, {\rm G}.
\label{BL}
\end{equation}
This value is also obtained from the energy ralation (\ref{W_tot})~\cite{ufn, Zaman}. 
In addition, we assume that the poloidal magnetic field remains regular (non-turbulent) throughout the jet, at least in its axial part. As already noted, this assumption is supported by both observations (the jets have a fairly regular polarization structure~\cite{Sokol, OGab}) and the results of numerical simulations~\cite{McTchB}.

Finally, for the so-called multiplicity factor $\lambda$, we assume
\begin{equation}
\lambda \sim 10^{11} - 10^{12}.
\label{lambda}
\end{equation}
This estimate follows from relation~\cite{ufn}
\begin{equation}
\lambda \sigma_{\rm M} \approx (W_{\rm tot}/W_{\rm A})^{1/2},
\label{lambdasigma}
\end{equation}
where \mbox{$W_{\rm A} = m_{\rm e}^2 c^5/e^2 \approx 10^{17}$ erg s$^{-1}$}, and $\sigma_{\rm M}$ is the so-called Michel magnetization parameter~\cite{Michel69}, which represents the maximum achievable hydrodynamic Lorentz factor in a strongly magnetized wind. The exact definition of this parameter is given below. Since both the values of superluminal velocities and the more detailed studies~\cite{Lenalambda} based on the core shift yield values of $\sigma_{\rm M} \sim 10$--$10^2$, we are inevitably led to the estimate (\ref{lambda}). The number density of the outflowing plasma $n_{\rm e}$ obtained for these values of $\lambda$ corresponds to the values previously obtained in~\cite{Lobanov}. It is these particles that determine the comoving reference frame.

Having determined the plasma parameters, we can now formulate the problem under consideration. As in most studies, we assume that at a given distance ${\bf r} = {\bf r}_{\rm inj}$ ($r_{\rm inj} \sim R_{\rm L}$) due to currently unknown processes in the comoving coordinate system, which moves with a Lorentz factor $\Gamma_{\rm inj}({\bf r})$ with respect to the laboratory frame, a power-law, isotropic distribution of radiating particles,
\begin{equation}
f(\gamma^{\prime}, \mu^{\prime}, {\bf r}_{\rm inj}) = \frac{n_{0}^{\rm com}}{2} \, f_{\gamma}(\gamma^{\prime}),  
\label{f}
\end{equation}
is formed. Here $n_{0}^{\rm com} = n_{\gamma}^{\rm com}(r_{\rm inj})$ represents the number density of particles in the comoving reference frame,
\begin{equation}
f_{\gamma}(\gamma^{\prime}) = A_{0}(\gamma^{\prime})^{-p},  \qquad \gamma_{\rm min} < \gamma^{\prime} < \gamma_{\rm max},  
\label{fgamma}
\end{equation}
and the normalization constant $A_{0}(\gamma_{\rm min}, \gamma_{\rm max})$ is determined by relation (\ref{norm}), where $\int f_{\gamma}(\gamma)\,{\rm d}\gamma = 1$:
\begin{equation}
A_{0}(\gamma_{\rm min}, \gamma_{\rm max}) = 
\frac{p-1}{\gamma_{\rm min}^{1-p} - \gamma_{\rm max}^{1-p}} \approx \frac{p-1}{\gamma_{\rm min}^{1-p}} .
\label{A0}
\end{equation}
As will be demonstrated below, the particle energy distribution function $f_{\gamma}(\gamma)$  can always be normalized to unity. Here and below we neglect the influence of the upper limit of the energy spectrum, assuming that \mbox{$\gamma_{\rm min} \ll \gamma_{\rm max}$.}
At the same time, the evolution of the power-law tail of radiating particles is considered separately of the evolution of the main (hydrodynamic) part of the jet.

In connection with the values of Lorentz factors for particles $\gamma \sim 10^2-10^{5}$ 
mentioned above,  it is necessary to make one more important remark. This concerns to the assumption applied in this paper regarding the absence of energy losses of radiating particles. It is clear that this approximation is only valid if the synchrotron emission time
$\tau \sim c/(\omega_{B}^2 r_{\rm e}\gamma)$ ($r_{\rm e} = e^2/m_{\rm e}c^2$ is the classical electron radius) is much longer than the escaping time of the particles from the region with a strong magnetic field, which is given by $t = r/c$. A similar estimation applies to Compton losses, which are typically of comparable magnitude to synchrotron losses. Consequently, owing to the quadratic decline of the magnetic field in the comoving reference frame (see below), it can be inferred that this approximation holds for distances $r$, satisfying the condition
\begin{equation}
\frac{r}{R_{\rm L}} > 10 
\left(\frac{\gamma}{10^{5}}\right)^{1/3}
\left(\frac{B_{\rm L}}{10 \, {\rm G}}\right)^{2/3}.
\label{rcr}
\end{equation}
It is precisely these distances that we apply further.

Let us now consider particles with positive radial velocity that move freely in the electromagnetic fields of a relativistic jet at distances $r > r_{\rm inj}$. At the same time, as demonstrated in~\cite{paper1}, the first adiabatic invariant,
\begin{equation}
I = \frac{(p_{\perp}^{\prime})^2}{h},
\label{first}
\end{equation}
is preserved with high accuracy.
Here $p_{\perp}^{\prime}$ is momentum perpendicular to the magnetic field in the comoving reference frame, and
\begin{equation}
h = \sqrt{{\bf B}^2 - {\bf E}^2}
\label{hcom}
\end{equation}
is the reduced magnetic field. As for the particles moving in the laboratory reference frame towards the central engine, their low number and, as will be demonstrated, relatively low energy allow them to be rejected painlessly.

It is worth noting that, given our interest in high-energy particles with $\gamma^{\prime} > \gamma_{\rm min}$, and since the preservation of the first adiabatic invariant, as we shall demonstrate, results in a reduction of their energy in the comoving reference frame, the lower limit imposed on the spectrum of radiating particles does not restrict the level of generality. We must keep in mind that, as previously stated, the number density $n_{\gamma}$  refers to the number of particles whose initial energy exceeded $m_{\rm e}c^{2}\,\gamma_{\rm min}$.

In conclusion, let us consider two more important points regarding the hydrodynamic Lorentz factor $\Gamma_{\rm inj}({\bf r})$, 
which will be of critical significance in the subsequent discussion. First, as is widely known~\cite{ufn, book, 
TchNar},  in the asymptotically far region outside the light cylinder, where $r_{\perp} \gg R_{\rm L}$ ($x \gg 1$) 
the hydrodynamic velocity approaches the drift velocity, $U_{\rm dr} = c \, [{\bf E}\times {\bf B}]/B^2$, and for 
values $x \ll \sigma_{\rm M}$ there is a straightforward asymptotic relation $\Gamma = x$. Here, and throughout
this paper, we use the convenient dimensionless coordinate $x = r_{\perp}/R_{\rm L}$. 

As for the region within the light cylinder, there is no consensus on the magnitude of the longitudinal velocity along the jet axis. There are arguments both in favor of the fact that only electrodynamic acceleration of particles is possible (and then the velocity on the axis is exactly zero~\cite{Kosh, Tomi, Toma}), and in favor of the fact that during the birth of particles, the outflowing plasma as a whole acquires a significant velocity $v \sim c$~\cite{BIP, CCPhPD}. Therefore, below we assume
\begin{equation}
\Gamma_{\rm inj} = (\Gamma_{0}^2 + x^2)^{1/2},
\label{Ginj}
\end{equation}
where $\Gamma_{0} \approx 1$ is a free parameter.

Secondly, we extensively employ the drift reference frame, rather than the comoving one. In fact, at $\Gamma_{0} > 1$ hydrodynamic motion consist of electric drift and the motion along magnetic field lines. This additional motion, as previously noted, becomes negligible outside the light cylinder; however, it plays a crucial role near the jet axis. Moreover, in general case, it is rather difficult to determine this additional velocity at distances $r \gg r_{\rm inj}$~\cite{Bogovalov}. On the other hand, to calculate the radiation (and it is for this main task that the present consideration is carried out), we can choose any reference frame in which the electric field is zero. In this regard, the drift reference frame seems to be the most suitable, since its determination requires knowledge only of the values of the electric and magnetic fields at a given location. Accordingly, only the values of the electric and magnetic fields at a given point will determine the Doppler factor $D = 1/(\Gamma_{\rm min}(1 - {\boldsymbol\beta}_{\rm min}{\bf n}))$ necessary to determine synchrotron radiation~\cite{Istomin, LPB}.
 
\section*{Monopole field --- exactly solvable problem}
\noindent

As an example, let us consider a model, but exactly solvable problem of a nonrotating conical jet in which there is only a monopole magnetic field
\begin{equation}
{\bf B}(r) = B_{0} \frac{R^2}{r^2} {\bf e}_{r},
\label{B}
\end{equation}
and $h = B$. The hydrodynamic flow occurs along the radial direction ${\bf e}_r$  with a constant velocity and a constant Lorentz factor $\Gamma=\Gamma_{\rm inj}$. 
This case models well the motion of particles within a light cylinder $x < 1$. At the same time, for the drift reference frame, we can apply $\Gamma_{\rm com} = 1$, that is, in this scenario, the drift reference frame coincides with the laboratory reference frame.

Consequently, the first adiabatic invariant remains constant in both the laboratory (drift) reference frame and the comoving reference frame (\ref{first}). Moreover, the total energy of particles also remains constant in the laboratory frame. This allows us to apply the known Lorentz transformations for each particle from the isotropic distribution (\ref{f}) in the comoving reference frame at $r = r_{\rm inj}$ to switch to the laboratory frame, and then, applying energy conservation and adiabatic invariant, to find the transverse momentum at an arbitrary point $r$. Thus, the final values of $\gamma$ and $\mu$ are determined.
Returning now to the comoving reference frame ($\Gamma = \Gamma_{\rm inj}$), or remaining in the laboratory reference frame ($\Gamma = 1$), it is not difficult to determine the two-dimensional distribution function $f(\gamma, \mu)$ at any distance of $r$ by counting the number of particles in the corresponding intervals $\Delta\gamma\,\Delta\mu$. It is only necessary to take into account that when determining the distribution function $f(\gamma,\mu)$ through conservation of the number of particles $f(\gamma,\mu)\Delta\gamma\Delta\mu = f(\gamma, ^{\prime}\mu^{\prime})\Delta\gamma^{\prime}\Delta\mu^{\prime}$, it is necessary to multiply it by the corresponding Jacobians of the transformation.

\begin{figure*}%[!ht]	
	\begin{minipage}{0.48\linewidth}
		\center{\includegraphics[width=1\linewidth]{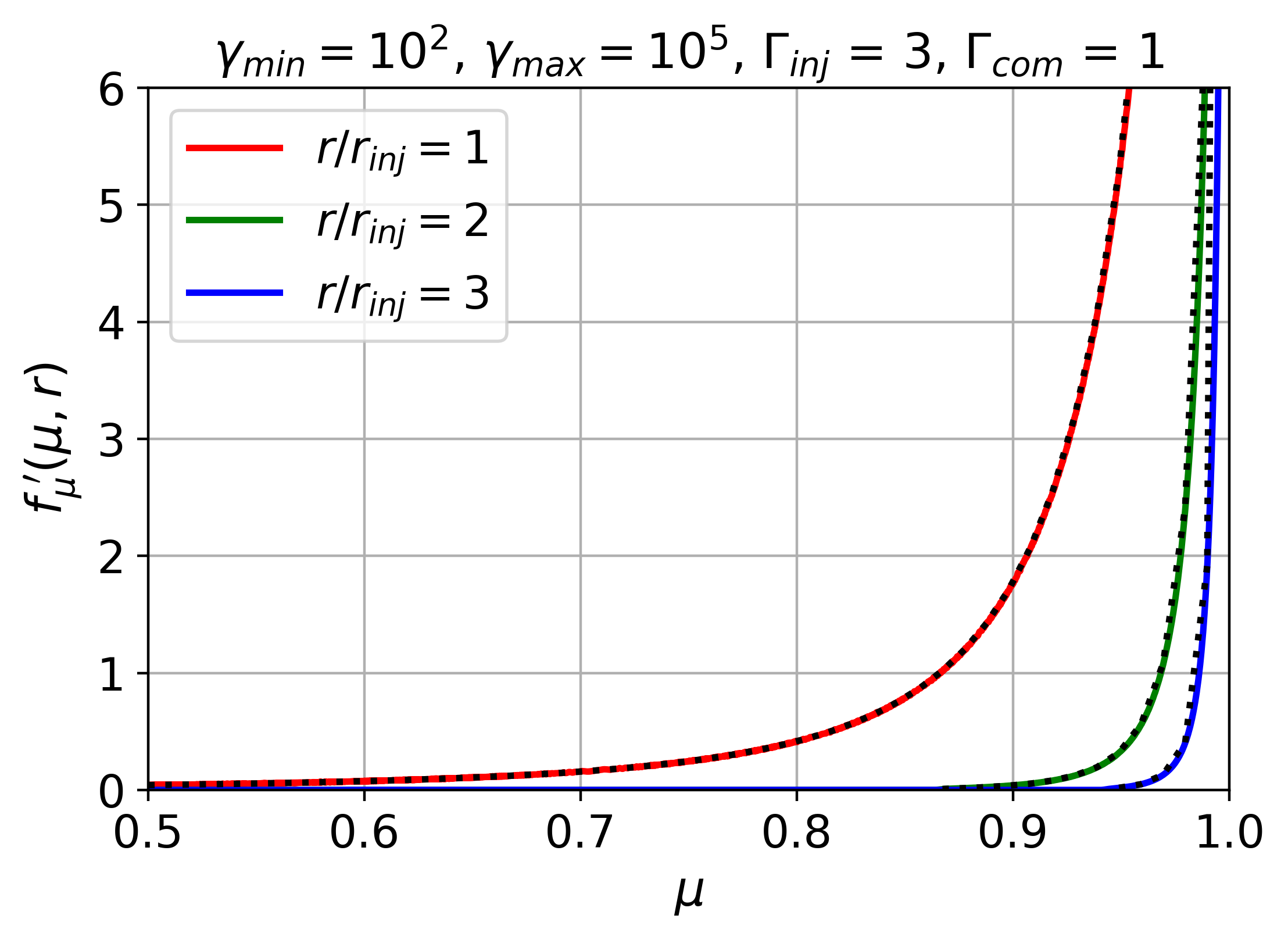} }
	\end{minipage}
	\hfill
	\begin{minipage}{0.48\linewidth}
		\center{\includegraphics[width=1\linewidth]{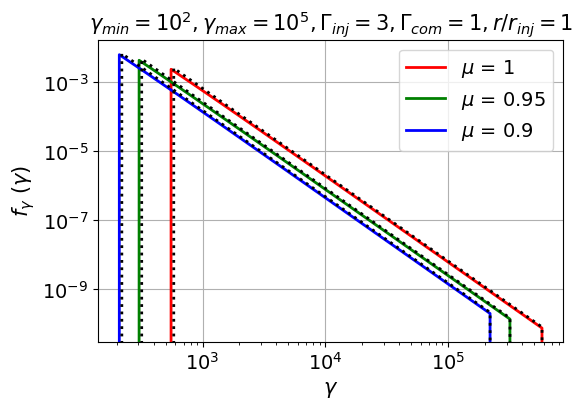}  }
	\end{minipage}
	\caption{On the left: normalized angular distributions $f^{\prime}_{\mu}(\mu, r) = n_{0}^{-1}\int f(\gamma, \mu, r){\rm d}\gamma$, where $\mu = \cos{\chi}$, $\chi$ is the pitch-angle, for various distances $r$ from the central engine in a drift (laboratory) reference frame  (\mbox{$\Gamma_{\rm inj} = 3$,} \mbox{$\Gamma_{\rm com} = 1$).} The dotted lines show the exact expression of the distribution function $f^{\prime}_{\mu}(\mu, r)$ (\ref{mudr}). On the right: the energy spectrum at a distance of $r =r_{\rm inj}$ for different values of the parameter $\mu$. The dotted lines show the exact expressions of the distribution function $f_{\gamma}(\gamma)$.}
\label{figure1}	
\end{figure*}

In figure~\ref{figure1} we show the distribution functions $f^{\prime}_{\mu}(\mu, r) = n_{0}^{-1}\int f(\gamma, \mu, r)\,{\rm d}\gamma$, normalized to the initial number density in the laboratory frame $n_{0} = n_{\gamma}(r_{\rm inj})$, describing the angular anisotropy of particles for different distances $r$ from the central engine, as well as the energy distributions $f_{\gamma}(\gamma,\mu, r)$ for different values of $\mu$ at $r=r_{\rm inj}$ in the drift reference frame ($\Gamma_{\rm inj} = 3$, $\Gamma_{\rm com} = 1$). Here and further in numerical calculations, we assume $\gamma_{\rm min} = 10^{2}$,  $\gamma_{\rm max} = 10^{5}$, and $p = 2.5$. These values, often used to explain the observed power-law spectra of jets in the radio range~\cite{pow2}, guarantee the fulfillment of both condition (\ref{rcr}), and condition $\gamma\gg 1$ with decreasing particle energy in the comoving reference frame.

As one can see, in this scenario, even at the initial point $r = r_{\rm inj}$, the angular distribution in the comoving reference frame has a strong shift towards small pitch angles (i.e., $\mu \rightarrow 1$). And at large distances the distribution very quickly becomes almost one-dimensional. For this reason, in Fig.~\ref{figure1} energy spectra are only shown for values of $\mu$ near unity; for lower $\mu$ (larger pitch angles) no radiating particles exist. As one can see, the energy spectrum maintains a power-law form in this coordinate system. Under these conditions, the number density $n_{\gamma}(r) = \int f_{\mu}(\mu, r)\,{\rm d}\mu$ approximately satisfies the relationships $n_{\gamma}(r) \propto r^{-2}$ (1.0, 0.24, and 0.11 corresponding to dimensionless densities $n_{\gamma}(r)/n_{0}$ for $r/r_{\rm inj} = 1$, 2, and 3, respectively). This is because in the laboratory reference frame, almost all particles moves in the direction of the magnetic field.

It is clear that in such geometry the problem can be solved analytically. Indeed, using the Lorentz transformations for ultrarelativistic particles,
\begin{eqnarray}
\gamma^{\prime} & = & \Gamma(1 - \beta\mu) \gamma, \\
\mu^{\prime} & = & \frac{\mu - \beta}{1 - \beta \mu}, 
\label{Lore}
\end{eqnarray}
where $\Gamma = \Gamma_{\rm inj}$, $\beta^2 = 1 - 1/\Gamma^{2}$, and the prime refers to the comoving reference frame, we obtain for the Jacobian of the transformation
\begin{equation}
\frac{\partial(\gamma^{\prime}, \mu^{\prime})}{\partial(\gamma, \mu)} = \frac{1}{\Gamma(1 - \beta\mu)}.
\label{Ja}
\end{equation}
Using now definition (\ref{dn2}) and the invariance of the distribution function $F(\mu, \gamma, r) = F(\mu^{\prime}, \gamma^{\prime}, r)$ for Lorentz transformations between laboratory and comoving reference frames, we obtain $f(\mu, \gamma, r)/\gamma^2 = f(\mu^{\prime},\gamma^{\prime}, r)/(\gamma^{\prime})^2$. As a result, in the laboratory reference frame  at $r = r_{\rm inj}$ we have
\begin{eqnarray}
{\rm d}n_{\gamma}(\gamma, \mu, r_{\rm inj}) = 
\frac{n_{0}}{2}\, \frac{{\rm d}\mu}{\Gamma^{4}(1 - \beta\mu)^{3}} \times \nonumber\\
\times \frac{(p-1)\gamma_{\rm min}^{p-1}}{\Gamma^{p-1} (1 - \beta\mu)^{p-1}} \,
\gamma^{-p} {\rm d}\gamma,
\label{fdrinj}
\end{eqnarray}
where $n_{0} = \int {\rm d}n_{\gamma}(\gamma, \mu, r_{\rm inj}) = \Gamma n_{0}^{\rm com}$ is the number density in the laboratory system at $r = r_{\rm inj}$.

Further, in the laboratory reference frame, due to the conservation of energy in the absence of an electric field, the Lorentz factors $\gamma$ of relativistic particles do not change during propagation from the injection radius $r_{\rm inj}$ to an arbitrary distance $r$. On the other hand, preserving the adiabatic invariant \eqref{first} in the laboratory reference frame for each particle leads to a change in the cosine of the particle pitch angle $\mu$ according to $(1-\mu^2)/|{\bf B}(r)| = {\rm const}$.  As a result, the distribution function $f(\mu,\gamma)$ at an arbitrary distance $r$ can be written in the following simple form
\begin{eqnarray}
{\rm d}n_{\gamma}(\gamma, \mu, r) = 
\frac{n_{0}}{2\Gamma}\, {\cal D}^{3}(\mu, r){\rm d}\mu \times \nonumber\\
\times (p-1)\Gamma_{\rm min}^{p-1}(\mu, r) \, \gamma^{-p} {\rm d}\gamma,
\label{fdr}
\end{eqnarray}
where the minimum Lorentz factor is
\begin{equation}
\Gamma_{\rm min}(\mu, r) = \frac{\gamma_{\rm min}}{\Gamma [1 - \beta \, {\cal M}(\mu, r)]},
\label{gammadr}
\end{equation} 
and
\begin{equation}
{\cal M}(\mu, r) = \left(1 - \frac{r^2}{r_{\rm inj}^2}(1 - \mu^2)\right)^{1/2}
\label{mur}
\end{equation}
is the cosine of the pitch angle that the relativistic particle had at $r_{\rm inj}$, and which is expressed in terms of the current pitch angle $\mu$ of the same particle using the conservations of energy and the adiabatic invariant~\eqref{first} in the laboratory reference frame. Accordingly, the Doppler factor has the form
\begin{equation}
{\cal D}(\mu, r) = \frac{1}{\Gamma [1 - \beta \, {\cal M}(\mu, r)]}.
\label{Ddr}
\end{equation} 

Thus, the distribution function of radiating particles still decomposes into the product 
of two distribution functions ${\rm d}n_{\gamma}(\gamma, \mu, r) = f_{\mu}(\mu ,r)f_{\gamma}(\gamma, \mu, r)  {\rm d}\gamma {\rm d}\mu$
with the same power-law energy distribution $f_{\gamma}(\gamma, \mu, r) \propto \gamma^{-p}$ (\ref{fgamma}) 
normalized to unity. 
However, as shown in Fig.~\ref{figure1}, the boundary Lorentz factor $\Gamma_{\rm min}(\mu, r)$ (\ref{gammadr}), and through it the normalization coefficient $A_{0}$, which is included in the definition of the function  $f_{\gamma}(\gamma, \mu, r)$ (\ref{fgamma}), depend on the $\mu$ parameter in the form
\begin{equation}
A_{0} = \frac{(p - 1)}{\Gamma^{1-p}_{\rm min}(\mu, r)}.
\label{A00}
\end{equation}
In this case, a decrease in the value of  $\Gamma_{\rm min}$, compared with $\gamma_{\rm min}$, means a decrease in the energy of particles in the comoving reference frame. %This occurs due to the action of the vortex electric field that appears in this system by the inhomogeneity of the magnetic field.
This is due to the presence of a vortex electric field, which is generated in this system as a result of the inhomogeneity of the magnetic field.

Hence, the energy distribution for ultrarelativistic particles, when relations (\ref{Lore}) are fulfilled, maintains its power-law form $f(\gamma, \mu, r) \propto  \gamma^{-p}$. Accordingly, for the angular distribution function with respect to the parameter $\mu$, we obtain for an arbitrary $r$
\begin{equation}
f_{\mu}(\mu, r) = \frac{n_{0}}{2} \, \frac{1}{\Gamma^{4}[1 - \beta \, {\cal M}(\mu, r)]^3}.
\label{mudr}
\end{equation}
The occurrence of the factor $g(\mu, r) = r^2(1-\mu^2)$ in relation (\ref{mur}) for ${\cal M}(\mu,r)$,  which is also included in the expressions (\ref{fdr}) and (\ref{gammadr}), is not accidental, since the distribution function must satisfy the kinetic equation~\cite{Roelof}
\begin{equation}
\mu \frac{\partial f}{\partial r} + \frac{1}{r} \, (1 - \mu^2) \frac{\partial f}{\partial \mu} 
= \frac{1}{c}\frac{\partial }{\partial \mu}\left(D_{\mu\mu}\frac{\partial f}{\partial \mu}\right),
\label{roelof}
\end{equation}
the solution of which, when neglecting angular diffusion (i.e., for $D_{\mu\mu} = 0$), is an arbitrary function of $g(\mu, r)$.

As a result, as shown in Fig.~\ref{figure1}, there is an excellent agreement between the exact analytical expression (\ref{mudr}) and the method we applied to determine the particle distribution functions. The same can be stated about the numerical values of the boundary Lorentz factors $\Gamma_{\rm min}(\mu, r)$. According to (\ref{gammadr}) we obtain $\Gamma_{\rm min} = 582$, $319$, and $220$ for the corresponding values $\mu = 1.0$, $0.95$, and $0.9$. 

Similarly, one can obtain the angular and energy distribution in the comoving reference frame ($\Gamma_{\rm com} = \Gamma_{\rm inj}$). It turns out that they can also be represented in the simple form of a product of two functions
\begin{eqnarray}
{\rm d}n_{\gamma}(\gamma^{\prime}, \mu^{\prime}, r) = 
\frac{n_{0}^{\rm com}}{2\Gamma^{3}}\, {\cal D}^{3}(\mu^{\prime}, r){\rm d}\mu^{\prime} \times \nonumber\\
\times (p-1)\Gamma_{\rm min}^{p-1}(\mu^{\prime}, r) \, (\gamma^{\prime})^{-p} {\rm d}\gamma^{\prime},
\label{fdrcom}
\end{eqnarray}
where again $n_{0}^{\rm com} = n_{\gamma}^{\rm com}(r_{\rm inj})$. On the other hand, the expression for $\Gamma_{\rm min}(\mu^{\prime}, r)$ 
\begin{equation}
\Gamma_{\rm min}(\mu^{\prime}, r)  = \frac{\gamma_{\rm min}}{\Gamma} \, {\cal D}(\mu^{\prime}, r), 
\label{Dcom}
\end{equation}
where the Doppler factor
\begin{eqnarray}
{\cal D}(\mu^{\prime}, r) = \Gamma^{-1} \Bigl[ 1 + \beta\mu^{\prime} - \beta((1 + \beta\mu^{\prime})^2 - \nonumber\\
 - (r^2/r_{\rm inj}^2)(1 - \beta^2)(1 - (\mu^{\prime})^2))^{1/2} \Bigr]^{-1},
\label{gamcom}
\end{eqnarray}
 is rather complex. However, as can be easily verified, this relation for $\Gamma_{\rm min}(\mu^{\prime}, r)$ satisfies the necessary conditions $\Gamma_{\rm min}(1, r) = \gamma_{\rm min}$ and \mbox{$\Gamma_{\rm min}(\mu^{\prime}, r_{\rm inj}) = \gamma_{\rm min}$}. Accordingly, ${\cal D}(\mu^{\prime}, r_{\rm inj}) = \Gamma$.

 \begin{figure*}	
	\begin{minipage}{0.48\linewidth}
		\center{\includegraphics[width=1\linewidth]{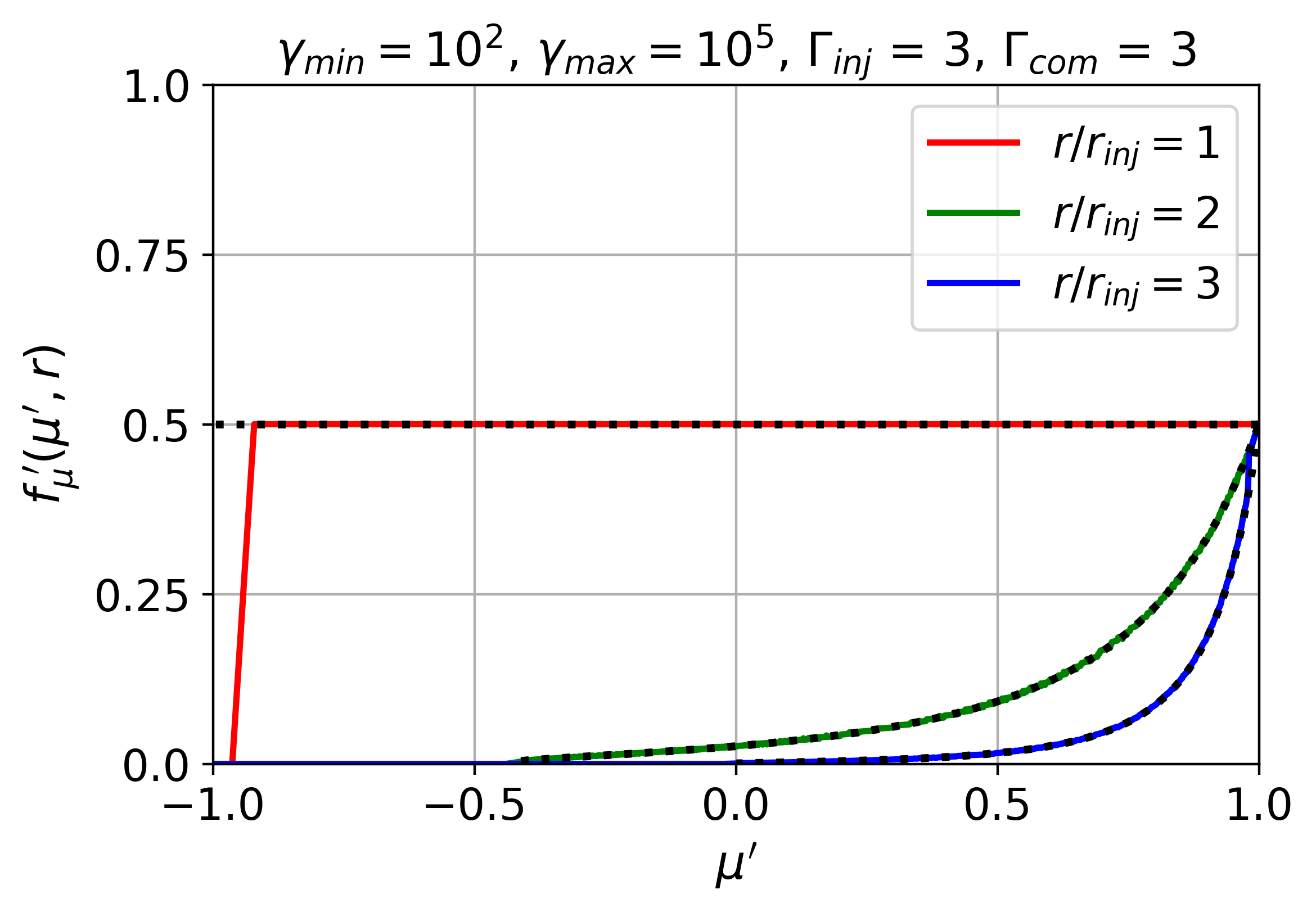} }
	\end{minipage}
	\hfill
	\begin{minipage}{0.48\linewidth}
		\center{\includegraphics[width=1\linewidth]{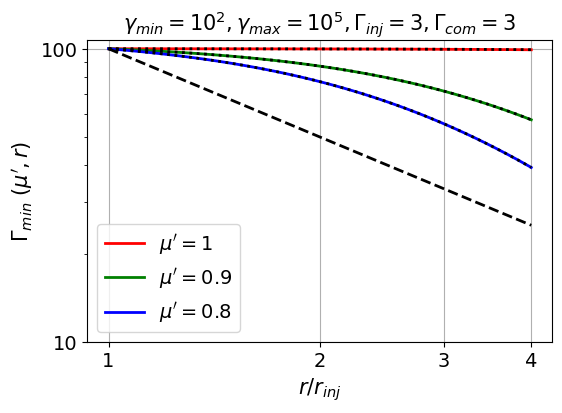}  }
	\end{minipage}
	\caption{On the left: angular distributions $f^{\prime}_{\mu}(\mu^{\prime}, r) = (n_{0}^{\rm com})^{-1}\int f(\gamma^{\prime}, \mu^{\prime}, r){\rm d}\gamma^{\prime}$ for various distances $r$ from the central engine in the comoving reference frame ($\Gamma_{\rm com} = \Gamma_{\rm inj} = 3$). On the right: the dependence of the minimum Lorentz factor  $\Gamma_{\rm min}(\mu^{\prime}, r)$  on the distance $r$ for different values of the parameter $\mu^{\prime}$. The dashed line shows the dependence of $\Gamma_{\rm min}(\mu^{\prime}, r) \propto h^{1/2} \propto r^{-1}$, applied in our previous paper~\cite{paper1}. The dotted lines show the obtained analytical dependencies (\ref{gamcom}).}
\label{figure2}	
\end{figure*}

As shown in Fig.~\ref{figure2}, an almost uniform distribution at $r = r_{\rm inj}$ is rapidly transformed into a narrowly directed distribution along the magnetic field. The absence of particles at $\mu^{\prime} \approx -1$, as we have already shown, is related to their motion towards the central engine. Based on the fact that particles with pitch angle cosines $\mu > 0$, are considered in the laboratory system, it is not difficult to obtain an expression for the minimum cosine of the pitch angle in the comoving reference frame at a finite distance along the magnetic field
\begin{equation}
\mu_{\rm min} = \frac{r/r_{\rm inj}\sqrt{r^2/r_{\rm inj}^2 - 1} 
- \Gamma_{\rm com}^2\beta}{\Gamma_{\rm com}^2\beta^2 + r^2/r_{\rm inj}^2}
\label{mumin}
\end{equation}
Simultaneously, the energy distribution once again maintains its power-law profile $f_{\gamma}(\gamma^{\prime}, \mu^{\prime}, r) \propto  (\gamma^{\prime})^{-p}$. 
With regard to the boundary values of $\Gamma_{\rm min}(\mu^{\prime}, r)$, as shown in Fig.~\ref{figure2}, the results of our calculation of the distribution function are in good agreement with the analytical dependencies $\Gamma_{\rm min}(\mu^{\prime}, r)$ (\ref{gamcom}). 

It is interesting to note that for the small distances shown in Fig.~\ref{figure2}, the number density $n_{\gamma}(r) = \int f(\mu, r)\,{\rm d}\mu$  does not satisfy the dependence $n_{\gamma}(r) \propto r^{-2}$ (1.0, 0.15, and 0.06 for a dimensionless number density $n_{\gamma}(r)/n_{0}^{\rm com}$ for $r/r_{\rm inj} = 1$, $2$, and $3$). This is due to the fact that at these distances not all particles move along the magnetic field, and therefore their average velocity along the magnetic field changes noticeably with increasing radius $r$, approaching the speed of light $c$ only at large distances.

The results obtained for such a simple model nevertheless allow us to formulate a number of significant differences identify several significant differences from the conventional model of an isotropic angular distribution. The key point here is not the strong angular anisotropy itself, which leads to a complete absence of particles at most pitch angles. As we demonstrate below, in real jets, this effect occurs only in the vicinity of the rotation axis. What is important here is the very fact of the dependence of energy spectra on the angular distribution, which was not previously taken into account.

Indeed, as has been shown, the normalization factor $A_{0} \propto \Gamma_{\rm min}^{p-1}$ (\ref{A00}) significantly depends on the parameter $\mu$. Therefore, for small pitch angles ($\mu \approx 1$), the minimum Lorentz factor $\Gamma_{\rm min}$, as shown in Fig.~\ref{figure2}, does not depend on distance, since for these particles, the contribution of changing transverse momentum to total particle energy is negligible. As pitch angles increase, $\Gamma_{\rm min} = \Gamma_{\rm min}(\mu^{\prime}, r)$ gradually approaches the asymptotic behavior $\Gamma_{\rm min} \propto B^{1/2} \propto r^{-1}$, shown in Fig.~\ref{figure2} by a dashed line, corresponding to the assumption applied in~\cite{paper1} that particle energy is completely determined by its transverse momentum. Clearly, this assumption is invalid for sufficiently small pitch angles. However, it should be noted that the energy spectrum index $p$ does not depend on the pitch angle and does not vary with distance.This conclusion is consistent with observations~\cite{Hovatta}.

Finally, the possibility of rejecting the contribution of particles moving against the main flow in a laboratory reference frame was confirmed.  This happens not only because at $\Gamma_{\rm com} > 1$ their relative number becomes small. More importantly, in the comoving reference frame they fall into the tails of the distribution both in terms of the value of $\mu$ and in terms of the energy $\gamma$.

\section*{Angular distribution of radiating particles in relativistic jets}
\noindent

Having checked the adequacy of our procedure for determining the distribution function for an exactly solvable problem, we proceed to our main task, i.e., to determine the angular and energy distribution of radiating particles in real relativistic jets. At the same time, we  examine their internal structure within a simple force-free approximation, which has been previously used in our research~\cite{paper1, MNRAS}
\begin{eqnarray}
{\bf B}_{\rm p} & = & \frac{\nabla \Psi \times {\bf e}_{\varphi}}{2 \pi r_{\perp}}, 
\label{Bp} \\ 
B_{\varphi} & = & - [1 + \varepsilon(\Psi)]\frac{\Omega_{\rm F}(\Psi)}{2\pi c} |\nabla \Psi|, 
\label{Bphi} \\ 
{\bf E}  & = & -  \frac{\Omega_{\rm F}(\Psi)}{2\pi c} \nabla \Psi,
\label{Er}
\end{eqnarray}
where the angular velocity $\Omega_{\rm F}(\Psi)$ is constant on magnetic surfaces $\Psi(r, \theta) =$ const. Additional correction of the force-free solution 
(\ref{Bphi})  by a small value $\varepsilon = \varepsilon(\Psi)$ is due to the fact that with this correction it is possible to reproduce one of the main properties of relativistic strongly magnetized flows already noted above, which consists in the fact that at large distances from the rotation axis, almost the total flux of electromagnetic energy is transferred into a  particle energy flux. As a result, at large distances from the jet axis, the energy of the hydrodynamic motion  $\Gamma mc^2$ along each magnetic surface $\Psi(r, \theta) =$ const becomes almost constant.

Indeed, since, regardless of the explicit form of the function $\Psi(r, \theta)$, the drift approximation is valid both in the region of strongly magnetized flow and in the saturation region, we obtain for fields (\ref{Bp})--(\ref{Er}) 
\begin{equation}
\Gamma_{\rm dr}^2  = \frac{1}{1 - U_{\rm dr}^2/c^2} = 
\frac{1 + (1 + \varepsilon)^2\omega^2 x^2}{1 + 2\varepsilon \omega^2 x^2 + \varepsilon^2 \omega^2 x^2},
\label{GUdr}
\end{equation}
where $U_{\rm dr} = c|{\bf E}|/|{\bf B}|$, $\omega(\Psi) = \Omega_{\rm F}(\Psi)/\Omega$, 
and again \mbox{$x = \Omega r_{\perp}/c$.} As a result, we have the asymptotic behavior $\Gamma_{\rm dr} = x$ ($\omega = 1$) already noted above for rather small distances from the jet axis $x < (2\varepsilon)^{-1/2}$, and $\Gamma_{\rm dr} = (2\varepsilon)^{-1/2}$ at large distances $x > (2\varepsilon)^{-1/2}$. Accordingly, the magnetic field in the drift reference frame \mbox{$h = \sqrt{{\bf B}^2 - {\bf E}^2}$} (\ref{hcom})
\begin{equation}
h \approx B_{\rm p} \sqrt{1 + 2 \varepsilon x^2}
\label{h}
\end{equation}
at small distances practically coincides with the poloidal magnetic field (and, therefore, falls as $r^{-2}$), whereas at large distances the condition $h \propto r^{-1}$ is fulfilled.

\begin{figure*}
	\begin{minipage}{0.48\linewidth}
		\center{\includegraphics[width=1\linewidth]{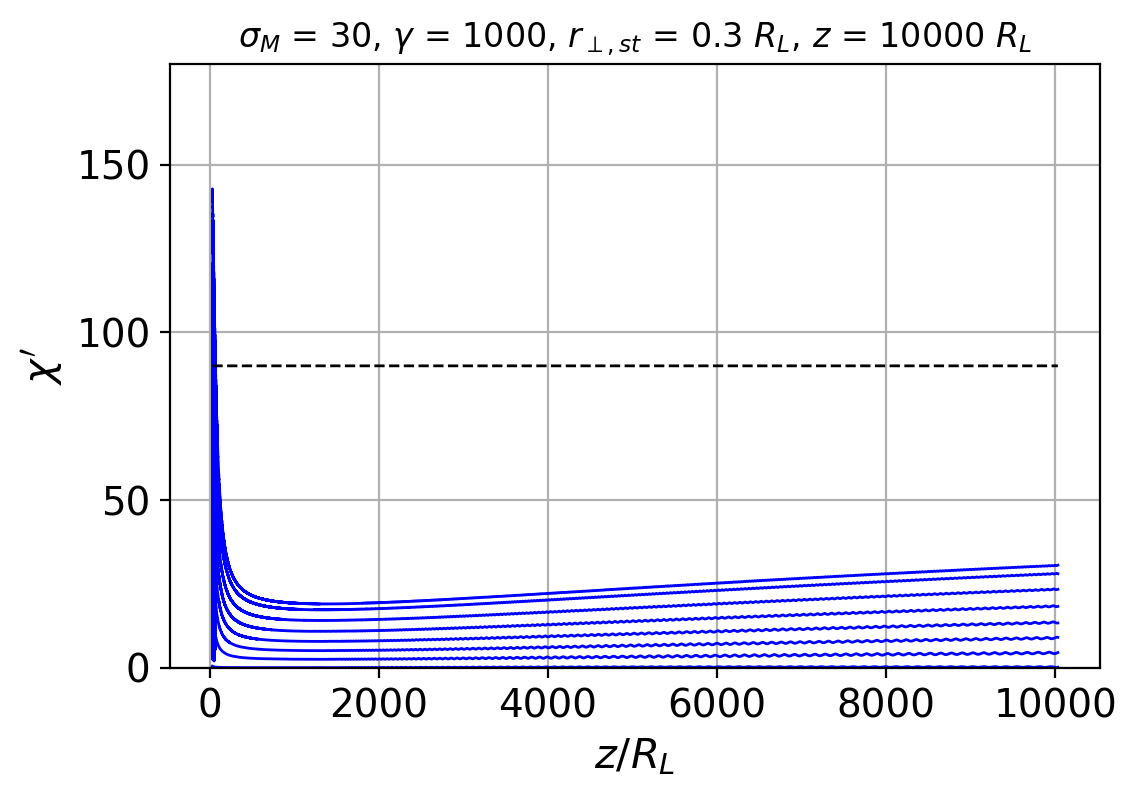} }
	\end{minipage}
	\hfill
	\begin{minipage}{0.48\linewidth}
		\center{\includegraphics[width=1\linewidth]{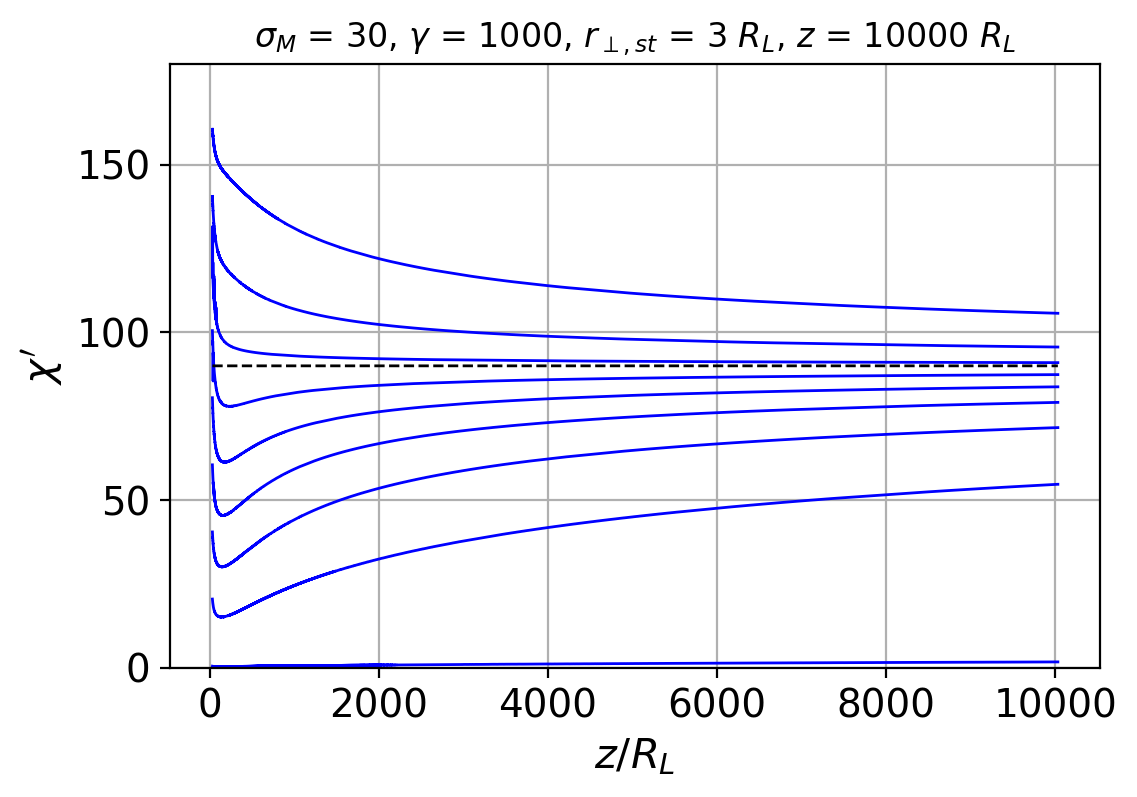}  }
	\end{minipage}
	\caption{Evolution of pitch angles for individual particles with initial Lorentz factors $\gamma=10^3$ in the comoving reference frame for the isotropic distribution of radiating particles at a distance of $z=30\, R_{\rm L}$ along the jet. The starting point is located inside ($r_{\perp, {\rm st}} = 0.3\,R_{\rm L}$, left) and outside ($r_{\perp, {\rm st}} = 3\, R_{\rm L}$, right) the light cylinder.}
\label{figure3}	
\end{figure*}

Finally, we note that the value $\varepsilon(\Psi)$ can be expressed in terms of the magnetization parameter $\sigma_{\rm M}$. The exact definition of the maximum Lorentz factor for the hydrodynamic flow, $\Gamma_{\rm max} = E(\Psi)/(\mu_{\rm e}\eta)$, can be derived from the Bernoulli equation~\cite{book}
\begin{equation}
E(\Psi) = \frac{\Omega_{\rm F}I}{2\pi} + \Gamma \mu_{\rm e}\eta,
\label{EPsi}
\end{equation}
where the first term corresponds to the electromagnetic energy flux, and the second one to the particles energy flux. Here $\mu_{\rm e}$ is the relativistic enthalpy, and $\eta$ is the ratio of the particle flux to the magnetic field flux, which we consider constant.  As one can see, the maximum Lorentz factor is different on different magnetic surfaces $\Psi =$ const depending on the value of $E(\Psi)$. On the other hand, assuming that the flow at the base of the jet is strongly magnetized, one can apply the solutions obtained for force-free configurations~\cite{Michel73, ff1}
\begin{equation}
E(\Psi) = \frac{\Omega_{\rm F}^{2}(\Psi)\Psi}{4\pi^2}.
\label{EPsiM}
\end{equation}
Therefore, it is convenient to express the value of $\varepsilon(\Psi)$ 
\begin{equation}
\varepsilon(\Psi) = \frac{1}{(1 - \Gamma_{\rm max}^{-2})^{1/2}}  - 1
\label{varep}
\end{equation}
in terms of the Michel magnetization parameter~\cite{Michel69}
\begin{equation}
\sigma_{\rm M} = \frac{\Omega^{2}\Psi_{\rm tot}}{8\pi^2\mu_{\rm e}\eta c^2},
\label{sigma}
\end{equation}
which represents the maximum Lorentz factor that can be achieved in relativistic jet with a full magnetic flux $\Psi_{\rm tot}$. As a result, we obtain
\begin{equation}
\Gamma_{\rm max}(\Psi) = \Gamma_{0} + 2 \omega^2(\Psi)\, \sigma_{\rm M}\, \frac{\Psi}{\Psi_{\rm tot}}.
\label{epsilon}
\end{equation}
Therefore, the condition for the termination of hydrodynamic acceleration $x = (2\varepsilon)^{-1/2}$ 
\mbox{($r = r_{\rm acc}$)} can be evaluated as
\begin{equation}
\frac{r_{\rm acc}}{R_{\rm L}} = \frac{\sigma_{\rm M}}{\theta_{\rm jet}},
\label{accel}
\end{equation}
where $\theta_{\rm jet}$ is the total angular half-width of the conical jet. Therefore, in the scenario under consideration, $r_{\rm acc}/R_{\rm L} > 100$, i.e., the plasma injection region with an isotropic particle distribution, $r_{\rm inj}/R_{\rm L} = 10$, occurs within the region where the hydrodynamic acceleration process is still ongoing.

With regard to the magnetic flux function, $\Psi(r, \theta)$, it should generally follow the so-called pulsar equation \cite{book}. However, in this paper we consider the simplest version of a thin conical jet $\Psi = \Psi_{0}(1 - \cos\theta)$ for $\theta < \theta_{\rm jet}$, 
which for $\theta_{\rm jet} \ll 1$ can be expressed as (see, e.g.,~\cite{Toma})
\begin{equation}
\Psi = \Psi_{\rm tot}\frac{\theta^{2}}{\theta_{\rm jet}^{2}}.
\label{Psi}
\end{equation}
Accordingly, we assume $\Omega_{\rm F} = \Omega$ ($\omega = 1$). However, general relations were given above, which make it possible to model a fairly wide class of relativistic jets.

\begin{figure*}	
	\begin{minipage}{0.48\linewidth}
		\center{\includegraphics[width=1\linewidth]{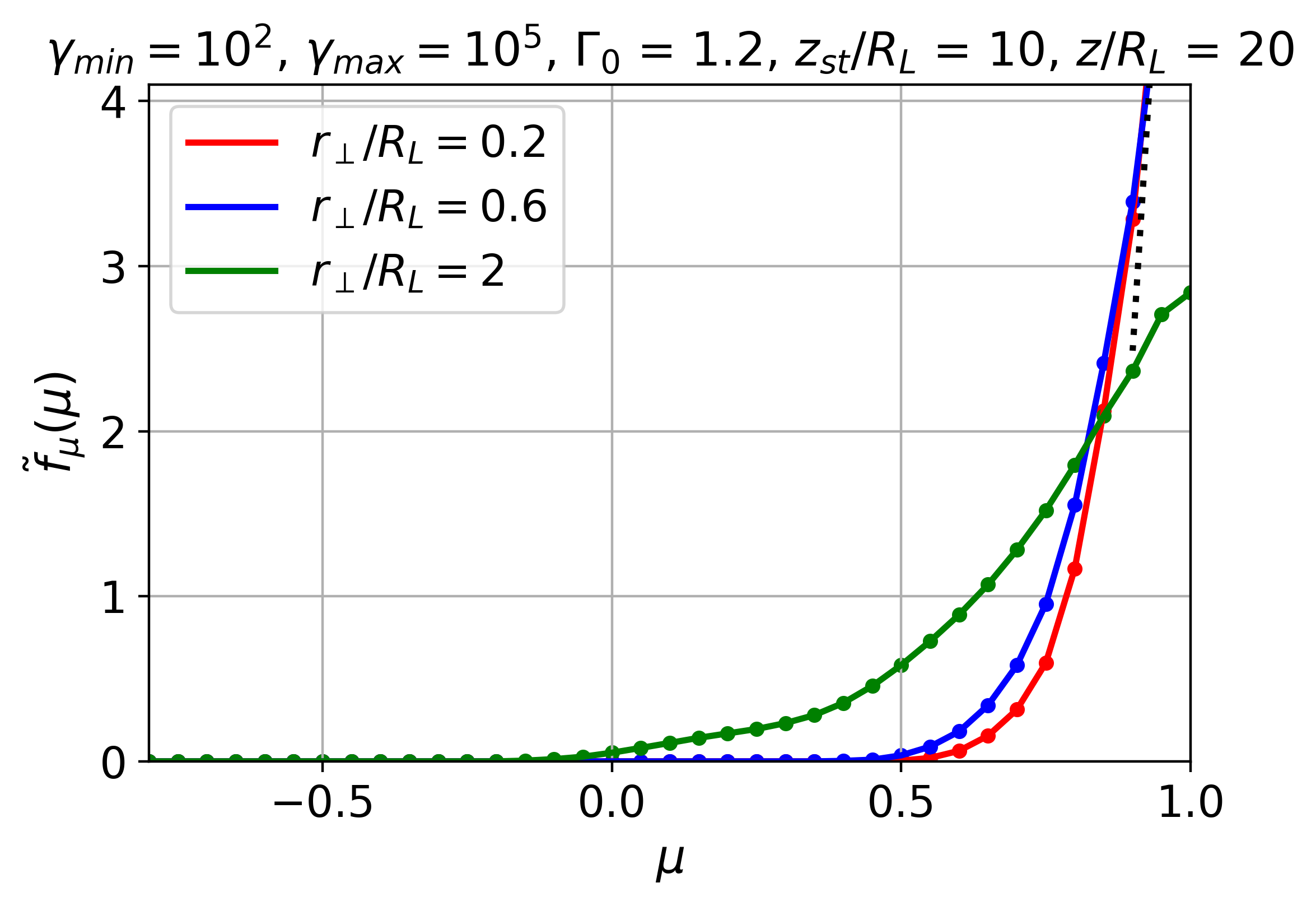} }
	\end{minipage}
	\hfill
	\begin{minipage}{0.48\linewidth}
		\center{\includegraphics[width=1\linewidth]{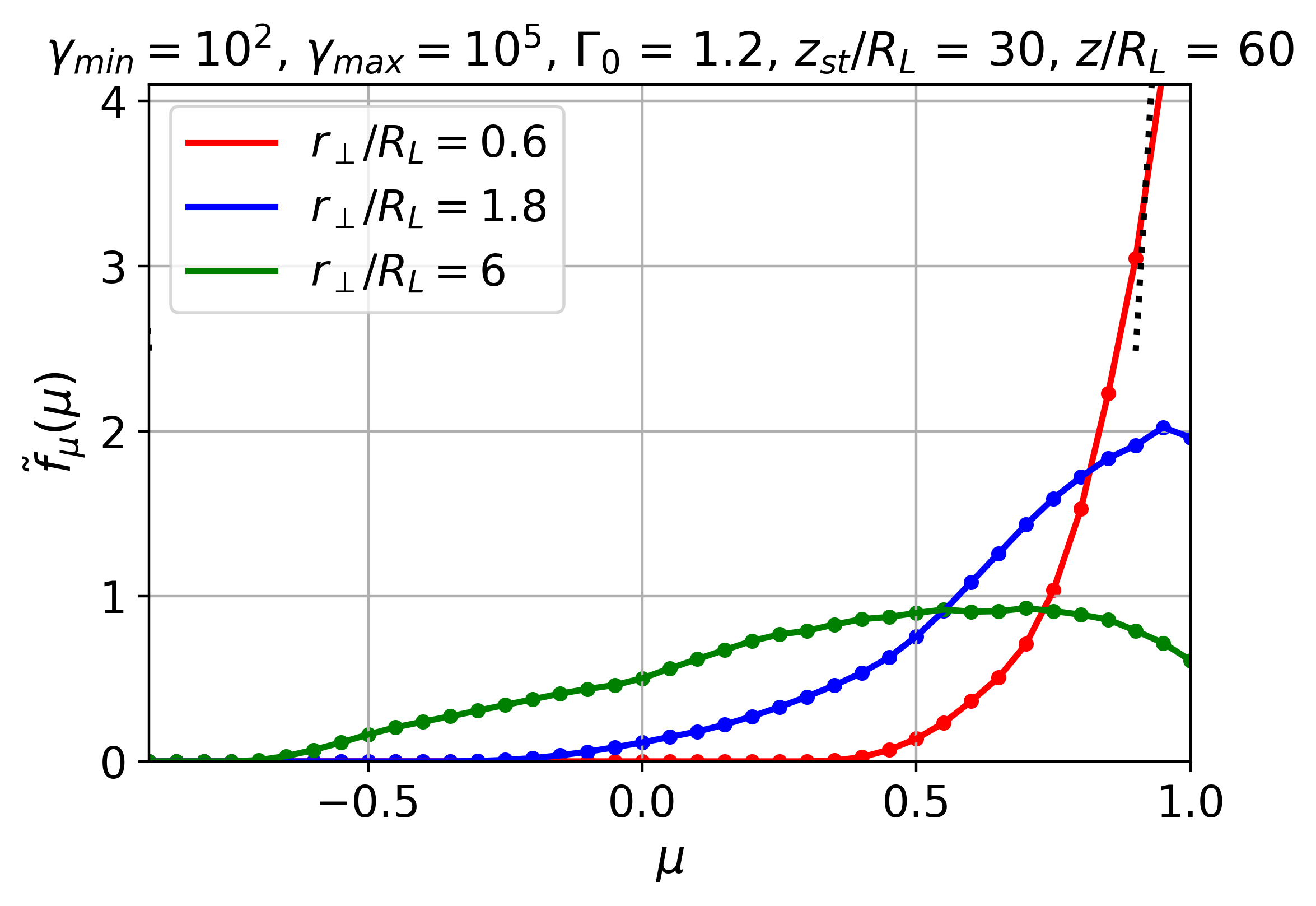}  }
	\end{minipage}
        \caption{Angular distributions ${\tilde f}_{\mu}(\mu) = n_{\gamma}^{-1}(z)\int f(\gamma, \mu){\rm d}\gamma$  for different starting points $z_{\rm st} = 10\, R_{\rm L}$ (left) and $z_{\rm st} =30\, R_{\rm L}$ (right) from the central engine at $z = 2\, z_{\rm st}$ for three trajectories with $r_{\perp}/z = 0.01$, $0.03$, and $0.1$. The dotted line shows the approximation (\ref{mudr}) that must be performed for all trajectories located within the light cylinder.} 
         \label{figure4}
\end{figure*}

\begin{figure*}
	\begin{minipage}{0.48\linewidth}
		\center{\includegraphics[width=1\linewidth]{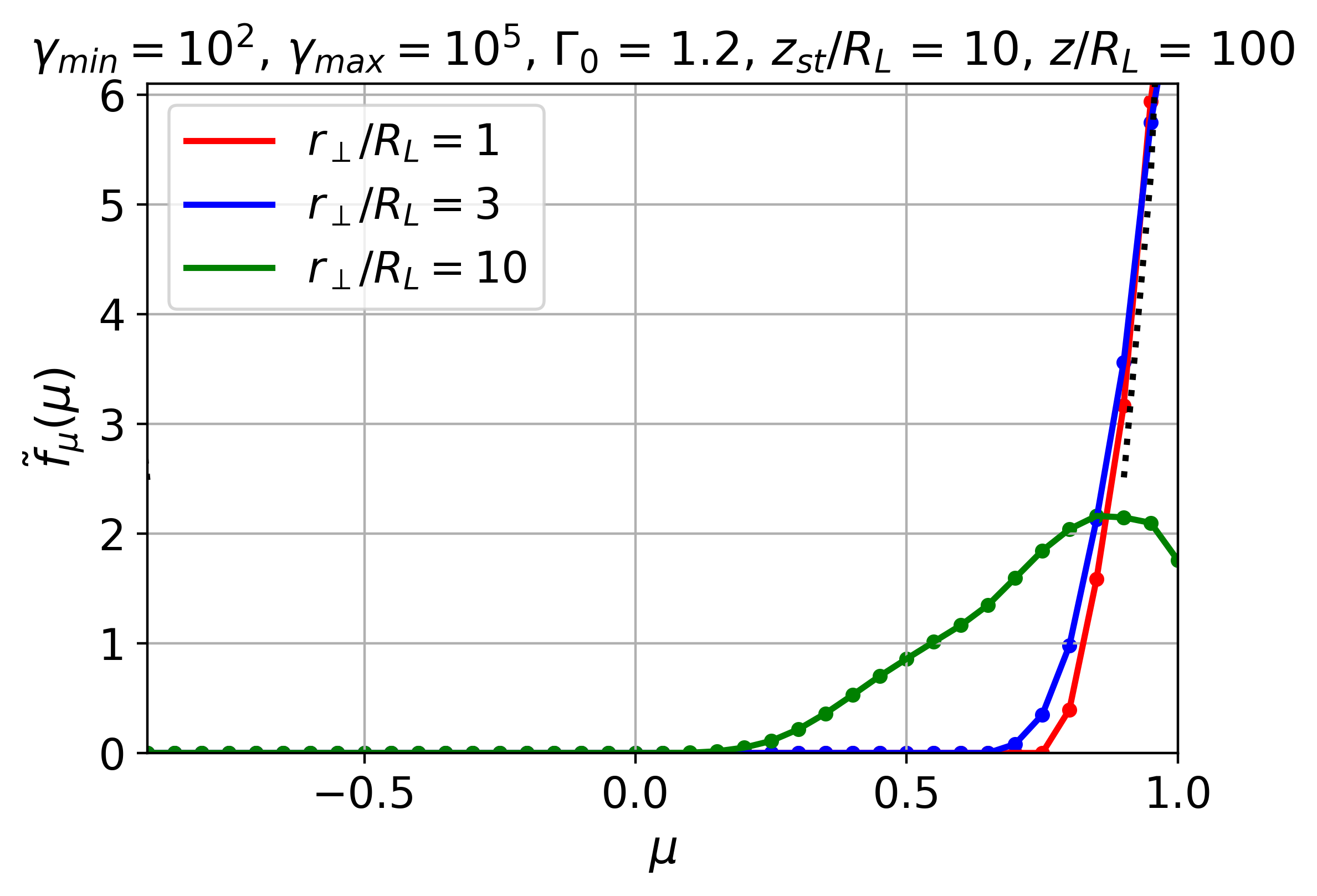} }
	\end{minipage}
	\hfill
	\begin{minipage}{0.48\linewidth}
		\center{\includegraphics[width=1\linewidth]{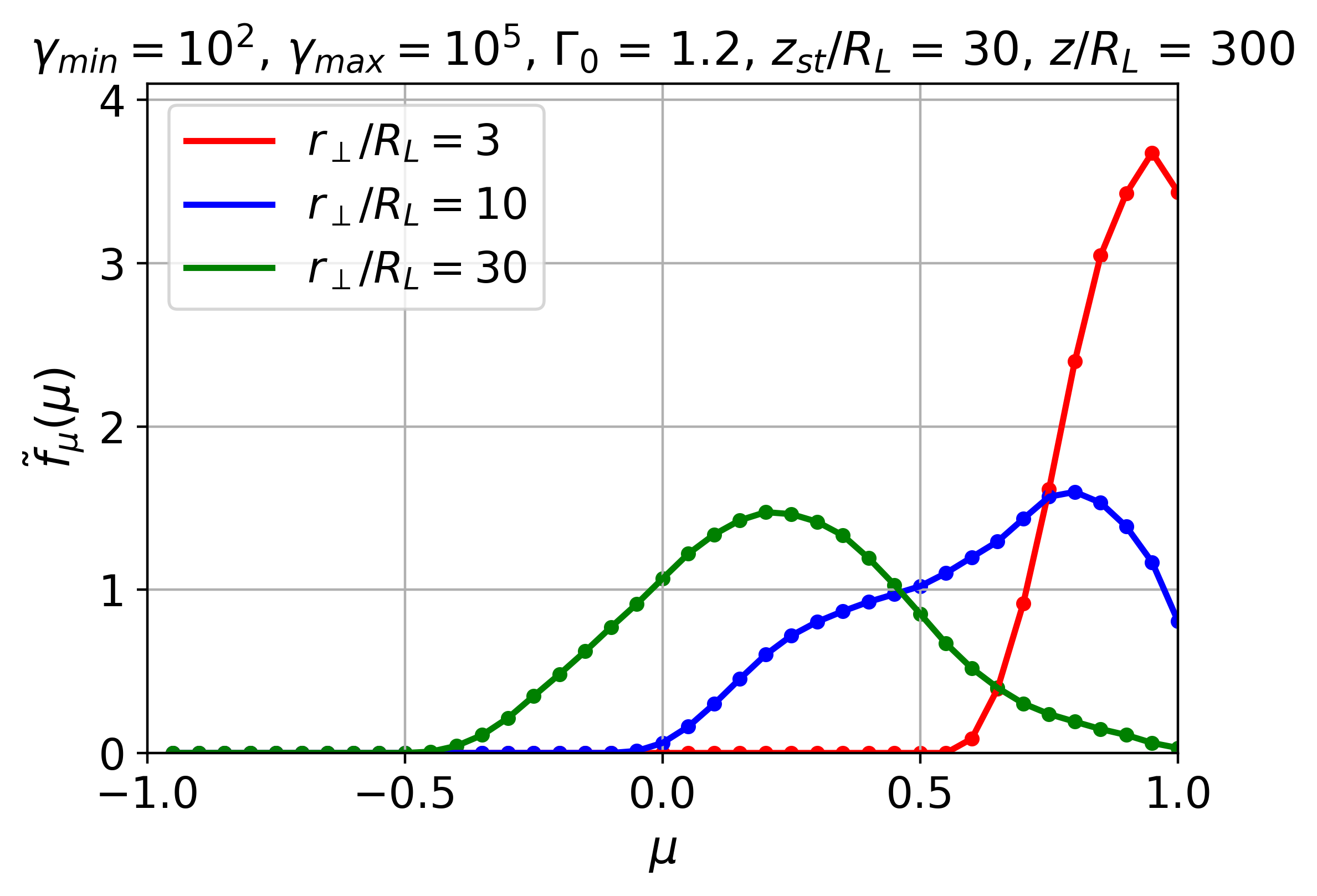}  }
	\end{minipage}
        \caption{The same for $z = 10\, z_{\rm st}$.} 
         \label{figure5}
\end{figure*}

As an example, in Fig.~\ref{figure3} the evolution of the pitch angles $\chi^{\prime}$ ($\mu^{\prime} = \cos\chi^{\prime}$) in the drift reference frame for the initially isotropic distribution of radiation particles with a Lorentz factor $\gamma = 10^3$ at the base of the relativistic jet at an altitude $z = 30 \, R_{\rm L}$ for $\sigma_{\rm M} = 30$ is shown. We choose starting points both within ($r_{\perp, {\rm st}} = 0.3 \,R_{\rm L}$) and outside ($r_{\perp, {\rm st}} = 3 \, R_{\rm L}$) the light cylinder. In this scenario, the value of $\Gamma_{\rm inj}$ is determined applying relation \mbox{$\Gamma_{\rm inj} = (\Gamma_{0}^2 + x^2)^{1/2}$ (\ref{Ginj})} for $\Gamma_{0} = 1.2$. 

Further, Fig.~\ref{figure4} shows the normalized angular distributions
${\tilde f}_{\mu}(\mu) = n_{\gamma}^{-1}(z)\int f(\gamma, \mu){\rm d}\gamma$  \mbox{($\int {\tilde f}_{\mu}(\mu){\rm d}\mu = 1$)} 
for different starting points \mbox{$z_{\rm st} = 10 \, R_{\rm L}$} (left) and \mbox{$z_{\rm st} = 30 \, R_{\rm L}$} (right)
for three trajectories with $(r_{\perp}/z) = 0.01$, $0.03$, and $0.1$ for $z = 2 \, z_{\rm st}$.
The dotted line on the left figure shows the approximation (\ref{mudr}),  which must be performed for all trajectories located within the light cylinder. Accordingly, Fig.~\ref{figure5} presents the results for $z = 10 \, z_{\rm st}$. Selection of distribution ${\tilde f}_{\mu}(\mu)$ is related to the fact that the distributions $f^{\prime}_{\mu}(\mu)$ defined in the previous section
satisfying the condition $\int f^{\prime}_{\mu}(\mu){\rm d}\mu = n_{\gamma}(z)/n_{0}$ fall rapidly decrease with increasing $z$. 

As we see, the evolution of pitch angles in the presence of an electric field differs significantly from their evolution in the conical jet without an electric field. Outside the light cylinder, where the electric field approaches the magnetic field in magnitude (and, therefore, the main motion becomes electric drift), the pitch angles in the drift reference frame approach $90^{\circ}$. On the other hand, as long as the trajectory of the particle is within the light cylinder, the pitch angles, as in the case discussed in the previous section, tend to $0^{\circ}$. Therefore, it is not surprising that their angular distribution turns out to be close to (\ref{mudr}) (dotted lines in Fig.~\ref{figure4}--\ref{figure5}).

The effect of orthogonalization of the angular distribution can be easily explained. To illustrate this, let us consider a simple Cartesian coordinate system ${\bf E} = E_0 \, {\bf e}_{x}$, ${\bf B} = [1 + \epsilon(z)] \, E_0 \, {\bf e}_{y}$, 
where $\epsilon(z) \rightarrow 0$ for $z \rightarrow \infty$ ($\epsilon(z) > 0$).
In these fields, the drift velocity of particles $U_{\rm dr} = c/(1 + \epsilon)$, directed along the $z$ axis, approaches the speed of light. Then the trajectory of any charge particle having initial pitch-angle $\chi^{\prime} = 90^{\circ}$ ($\mu^{\prime} = 0$)
will lie in the $xz$ plane, as it is not affected by a force in the $y$-direction. This implies that the pitch angle remains constant and equal to $90^{\circ}$.

Thus, we conclude that the angular distribution of radiating particles in relativistic jets should strongly depend on the distance from the central engine at which it ceases to be isotropic. If this happens at short distances ($z_{\rm st} < 10 R_{\rm L}$), when the trajectories of almost all particles begin within (or near) the light cylinder, then the angular distribution at large distancies remains strongly anisotropic in the direction of the magnetic field  (Fig.~\ref{figure3}, left). Otherwise, there should be orthogonal anisotropy, when the pitch angles tend to $90^{\circ}$ (Fig.~\ref{figure3}, right). At the same time, as shown in Fig.~\ref{figure4}--\ref{figure5}, the further away the particle trajectories are from the light cylinder, the more particles have pitch angles close to $90^{\circ}$.

\begin{figure*}
	\begin{minipage}{0.48\linewidth}
		\center{\includegraphics[width=1\linewidth]{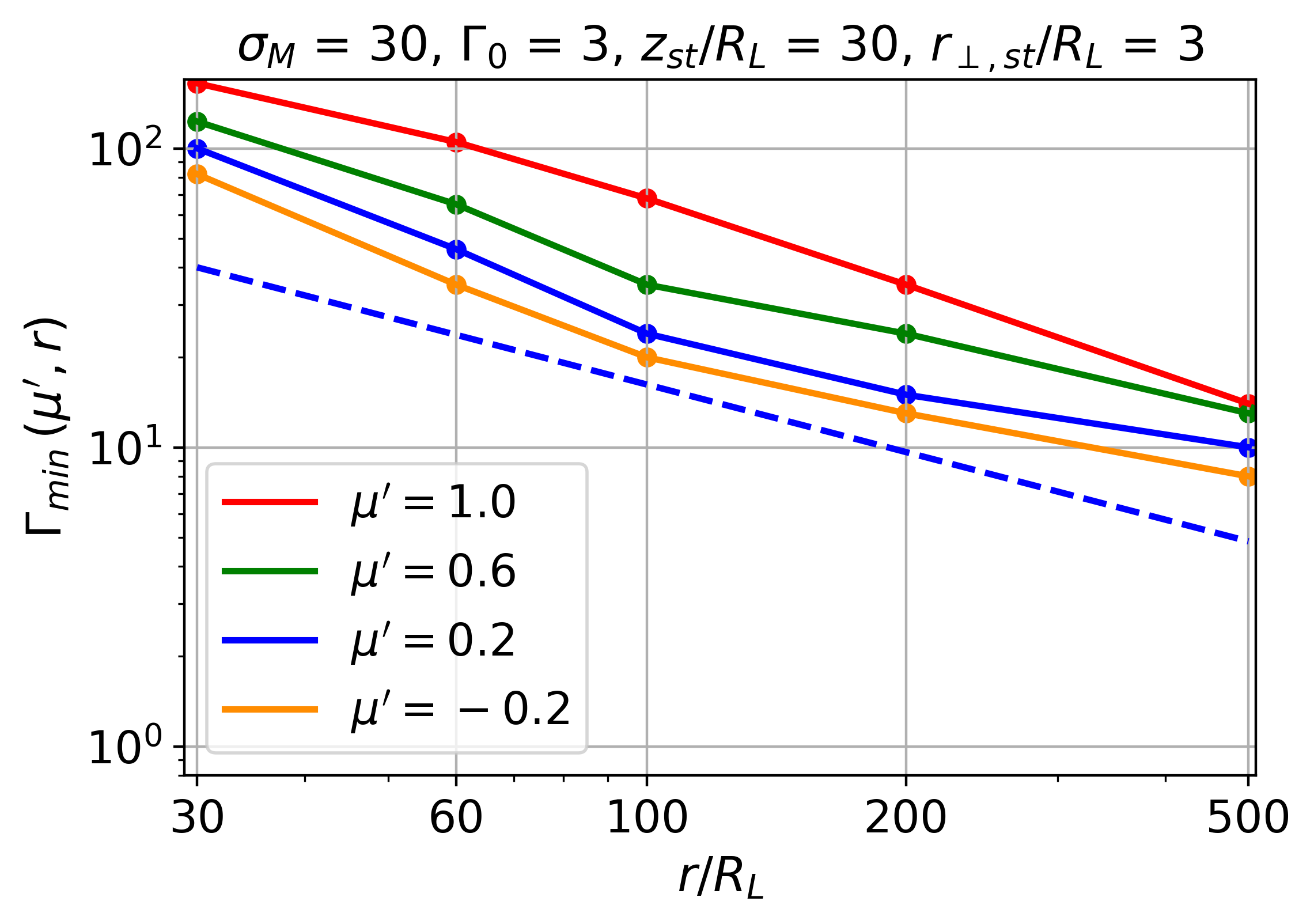} }
	\end{minipage} 
	\hfill
	\begin{minipage}{0.48\linewidth}
		\center{\includegraphics[width=1\linewidth]{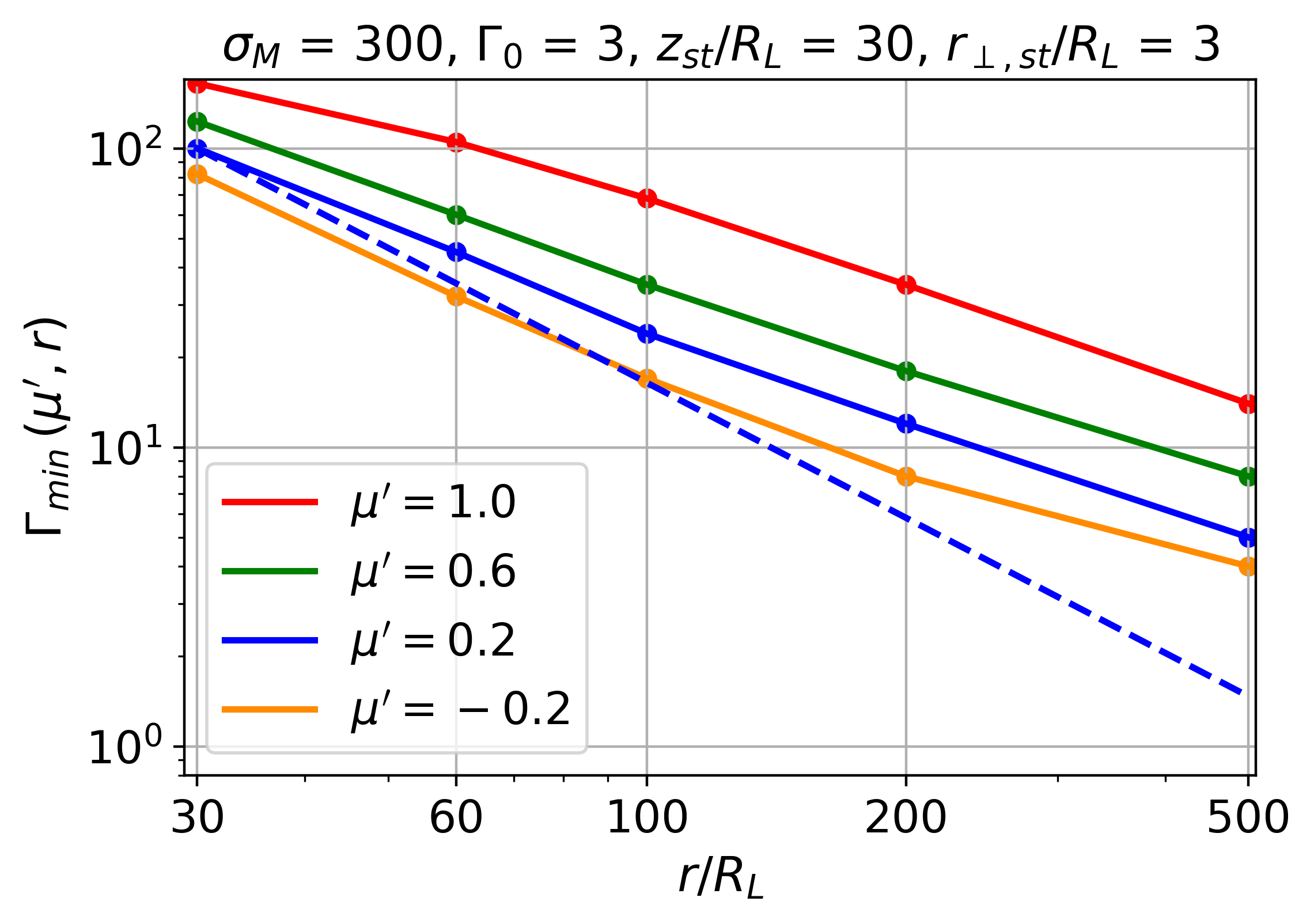}  }
	\end{minipage}
        \caption{Dependencies of the minimum Lorentz factors $\Gamma_{\rm min}(\mu^{\prime}, r)$ on the radius $r$ for $\sigma_{\rm M} = 30$ (left) and $\sigma_{\rm M} =300$ (right). On the left, the dashed line corresponds to the dependence $\Gamma_{\rm min} \propto h^{1/2} \propto r^{-1/2}$, which is valid after saturation, when the energy flux is associated with the hydrodynamic flow. On the right the dashed line corresponds to the dependence $\Gamma_{\rm min} \propto h^{1/2} \propto r^{-1}$, which is valid until saturation is reached.} 
         \label{figure6}
\end{figure*}

As for the spectrum of radiating particles, as was shown above, its evolution can be conveniently represented through a change in the minimum Lorentz factor $\Gamma_{\rm min}(\mu^{\prime}, r)$ (here, we store the primes, recalling that we are considering energy distributions not in the laboratory reference frame, but in the drift reference frame).  The corresponding dependencies are shown in Fig.~\ref{figure6}.

As one can see, both for small values of $\sigma_{\rm M}$ (Fig.~\ref{figure6}, left), when radiating particles almost immediately pass into the saturation region, when the energy flux is associated with the hydrodynamic flow, and for large values of $\sigma_{\rm M}$ (Fig.~\ref{figure6}, right), when the initial part of the trajectory lies in the region of a strongly magnetized flow, for the values of $\mu^{\prime}\sim 1$ ($\chi^{\prime}\sim 0^{\circ}$), relation $\Gamma_{\rm min} \propto r^{-1/2}$ is fulfilled with good accuracy. As in the case of the example discussed in the previous section, this is due to the fact that for $\chi^{\prime}\sim 0^{\circ}$ the main energy is contained in longitudinal motion, and therefore the preservation of the first adiabatic invariant does not affect the total energy of the emitting particles. However, for values of $\mu^{\prime}$ close to zero (i.e. for pitch angles close to $90^{\circ}$), this approximation does not take place at small distances.  At the same time, here, as in the previous case, the slope of the power-law energy spectrum is also preserved.

\section*{Discussion and Conclusion}
\noindent

Thus, it has been demonstrated that for relativistic flows in which the electric field is of primary importance, the conservation of the first adiabatic invariant does not result in a reduction of the pitch angles of radiating particles as they move into the region of weak magnetic field. This is due to the drift nature of the motion of particles outside the light cylinder. 

It is important to note the following points. First of all, let us recall that we discussed the particle distribution function in the drift reference frame. As was mentioned, this reference frame is particularly useful for determining the observed radiation, as the Doppler factor of the hydrodynamic flow $D$ can be easily determined.

Further, we emphasize the very possibility of factoring the distribution function $f(\gamma, \mu, {\bf r})$ into the product of two functions $f_{\mu}(\mu, {\bf r})$ and $f_{\gamma}(\gamma, \mu, {\bf r})$. At the same time, the distribution function of radiating particles over the pitch angles $f_{\mu}(\mu, {\bf r})$ does not depend on the energy of the particles, but significantly depends on the distance to the jet axis: near the axis, the distribution function is strongly shifted to the region of small pitch angles, whereas at large distances, on the contrary, most of radiating particles have pitch angles close to $90^{\circ}$.

Finally, regarding the energy distribution function $f_{\gamma}(\gamma, \mu, {\bf r})$, it should be noted that the slope of the power-law spectrum of radiating particles, $p$, is preserved as they propagate along the jet.  This, as noted earlier, is consistent with observations~\cite{Hovatta}.  In turn, this allows us to characterize the high-energy tail of the energy spectrum with only one function $\Gamma_{\rm min}(\mu^{\prime}, {\bf r})$, which, however, significantly depends on both the pitch angle $\chi^{\prime}$ and the distance from the jet axis. At the same time, for angles  $\chi^{\prime} \sim 90^{\circ}$ ($\mu^{\prime} \sim 0$) we obtain $\Gamma_{\rm min} \propto h^{1/2}$ (it is easily obtained from the assumption that all the energy of radiating particle is contained in its transverse motion~\cite{paper1}).

In conclusion, we would like to repeat that in this paper we deliberately refrained from including possible angular isotropization processes in our analysis. Our main goal was not to determine a specific angular distribution function of radiating particles, but to demonstrate that there is a permanent anisotropization mechanism in the relativistic wind, which acts not in the direction of small angles, as in the case of a pure magnetic field, but in the direction of angles close to 90 degrees. This is the main result of our  study.

Nevertheless, it is worthwhile to at least briefly discuss the relevance of the individual particles motion model, in which isotropization processes do not play a significant role. It should be noted that this simple model is not applicable to the main component of the plasma, which is typically described within the framework of a single-fluid model of magnetic hydrodynamics (MHD)~\cite{MHD1, MHD2}. As for the high-energy tail of interest, where two-particle collisions are rare and do not have a significant effect on the particle distribution function, isotropization processes for these particles are determined by the level of turbulence in the plasma, associated, e.g., with the generation of Alfv{\'e}n waves. This type of turbulence is well known~\cite{turb}, and occurs effectively when particle velocities exceed the Alfv{\'e}n velocity $V_{\rm A} = B/\sqrt{4 \pi \rho}$. However, it is evident that Alfv{\'e}n turbulence is inhibited in situations where the above nonrelativistic expression for $V_{\rm A}$ exceeds the speed of light, $c$. Rewriting now the weak turbulence condition, $V_{\rm A} > c$, in the form
\begin{equation}
\frac{r}{R_{\rm L}} < 100 
\left(\frac{\lambda_{\gamma}}{10^{10}}\right)^{-1/2}
\left(\frac{\Omega}{10^{-6} \, {\rm s}^{-1}}\right)^{-1/2}
\left(\frac{B_{\rm L}}{10^2 \, {\rm G}}\right)^{1/2},
\label{Alfven}
\end{equation}
we conclude that at least at the base of the jet, the isotropization processes associated with Alfv{\'e}n turbulence are strongly suppressed.

It is also necessary to mention the Weibel instability, which arises in weakly collisional plasma with anisotropic angular distribution and can make a significant contribution to the isotropization of the particle distribution function. At present \cite{Kocharovsky1}, there is no clear understanding of the dependence of the Weibel threshold, saturation, and damping on the shape of the particle distribution, even in the absence of an external magnetic field.  However, it is known that in a magnetized weakly collisional plasma, the Weibel instability will not occur at a certain strength of the external magnetic field \cite{Kocharovsky2}. To estimate, we assume that the maximum increment value occurs in the case of a strong anisotropy in the distribution function of radiating particles. This increment is on the order of the relativistic plasma frequency, $\omega_{\rm p}$, and the relativistic cyclotron frequency, $\omega_{\rm c}$. Then, the condition for the suppression of Weibel instability by an external magnetic field can be expressed as follows:
\begin{equation}
    \omega_{\rm c} \gtrsim \omega_{\rm p}.
\end{equation}
Hence, at least near the jet axis, we obtain distances
\begin{eqnarray}
\frac{r}{R_{\rm L}} \lesssim 100 \left(\frac{\gamma}{10^{2}}\right)^{-1/2}\left(\frac{\lambda_{\gamma}}{10^{10}}\right)^{-1/2} \times \nonumber\\
\times \left(\frac{\Omega}{10^{-6} \, {\rm s}^{-1}}\right)^{-1/2}
\left(\frac{B_{\rm L}}{10^2 \, {\rm G}}\right)^{1/2},
\end{eqnarray}
for which the isotropization process due to the Weibel instability does not occur. Thus, the possibility of using our approach in studying the evolution of the angular distribution within an expanding relativistic jet on the scales that have been explored in this study is confirmed.

Therefore, we can reasonably expect that for distances up to approximately 100 $R_{\rm L}$, the model of the free particle motion in a uniform magnetic field that we have applied is a reasonable approximation. For larger distances, as equation (\ref{roelof}) shows, the efficiency of isotropization depends on the diffusion coefficient $D_{\mu\mu}$, namely on the ratio \mbox{$\xi = c/(D_{\mu\mu}r)$}. Above, we actually considered the case of weak isotropization of $\xi \gg 1$ (strong focusing in a pure magnetic field according to the terminology of the paper~\cite{HeSchli}). As for the opposite case of $\xi\ll 1$, a detailed discussion of this issue, which requires  of the value $D_{\mu\mu}$, is beyond the scope of this paper. 

The authors thank to M.A. Garasev, V.V. Kocharovsky, A.P. Lobanov, and D.O. Chernyshov for their valuable discussions. This study was supported by the Russian Science Foundation, project no. 24-22-00120.

{}

\end{document}